\documentclass[]{raa}            
\usepackage{graphicx,times}
\usepackage{amssymb}
\usepackage{natbib}

\begin{document}
   \title{Asteroseismology of DAV white dwarf stars and G29-38
}

 \volnopage{ {\bf 0000} Vol.\ {\bf 0} No. {\bf XX}, 000--000}
   \setcounter{page}{1}

   \author{Y. H. Chen
      \inst{1, 2, 3}
   \and Y. Li
      \inst{1, 2}}

   \institute{\inst{1}  Yunnan Astronomical Observatory, Chinese Academy of Sciences, Kunming 650011, China; {yanhuichen1987@ynao.ac.cn, ly@ynao.ac.cn\\
   \inst{2}  Key Laboratory for the Structure and Evolution of Celestial Objects, Chinese Academy of Sciences, Kunming 650011, China\\
   \inst{3}  University of Chinese Academy of Sciences, Beijing 100049, China}
~~~{\it }\\
\vs \no
   {\small Received [0000] [July] [day]; accepted [0000] [month] [day] }
}

\abstract{Asteroseismology is a powerful tool to detect the inner structure of stars. It is also widely used to study white dwarfs. In this paper, we discuss the asteroseismology work of DAV stars. The detailed period to period fitting method is fully discussed, including the reliability to detect the inner structure of DAV stars. If we assume that all observed modes of some DAV star are the $l$ = 1 ones, the errors of model fitting will be always great. If we assume that the observed modes are composed of $l$ = 1 and 2 modes, the errors of model fitting will be small. However, there will be modes identified as $l$ = 2 without quintuplets observed. G29-38 has been observed spectroscopically and photometrically for many years. Thompson et al. (2008) made $l$ identifications for the star through limb darkening effect. With eleven known $l$ modes, we also do the asteroseismology work for G29-38, which reduces the blind $l$ fittings and is a fair choice. Unfortunately, our two best-fitting models are not in line with the previous atmospheric results. Based on factors of only a few modes observed, stability and identification of eigenmodes, identification of spherical degrees, construction of physical and realistic models and so on, detecting the inner structure of DAV stars by asteroseismology needs further development.
  }

   \authorrunning{Y. H. Chen \& Y. Li}            
   \titlerunning{Asteroseismology of DAV white dwarf stars and G29-38}  
   \maketitle


%
%

\section{Introduction}

White dwarfs are the final evolutionary stage of about 98\% of stars and some of them are $g$-mode pulsators (Winget \& Kepler 2008). Along the cooling curve of white dwarfs, there are DOV (GW Vir), DBV (V777 Her), and DAV (ZZ Ceti) instability strips. DA white dwarfs comprise about 80\% of all white dwarfs and DA white dwarf seismology is a focus of the white dwarf seismology research (Bischoff-Kim \& Metcalfe 2011). Asteroseismology is a powerful tool to analyze the inner structure of stars. By comparing grid-model calculations to the observation results, a best-fitting model will be selected which can basically reflect the inner structure of the observed star.

Long time and high signal-to-noise photometric observations are required to ensure an effective mode identification. For this work, there are sometimes combination frequencies, especially at the red edge of the instability strip where the convection is efficient. Selecting the "real modes" from combination frequencies is an important task, which is related to whether we are fitting eigenmodes. Basically, if there is no parent modes (A and B), the coincided daughter mode (C = A + B) will be considered as a real mode. However, if there is a relation of A + B = C for the three frequencies, we should decide which is the difference frequency. Each eigenmode is characterized by a group of indices ($k$, $l$, $m$). $k$ is the radial order, $l$ the spherical degree and $m$ the azimuthal order. The radial orders are impossible to determine from the observations, which describe the nodes of standing wave of the radial eigenfunction inside a star. The values of $l$ and $m$ can be constrained by observing rotational splits. Actually, if we observe triplets, the modes will be considered as $l$ = 1 modes. If we observe quintuplets, the modes will be considered as $l$ = 2 modes. For PG 1159 stars, the quintuplets have been observed and identified, such as PG 1159-035 (Costa et al. 2008). For GD358, there are also quintuplets observed (Provencal et al. 2009). However, as far as we know, there has no quintuplets observed for DAV stars, which may be caused by small amplitude and the variabilities of amplitude and frequency for quintuplet modes. While, the triplets can be observed in DAV stars, such as EC14012-1446 (Handler et al. 2008; Provencal et al. 2012).

Since DAV stars have low luminosity and low effective temperature, it is a challenge for aseroseismology with only a few modes observed. It is necessary to find a proper method to study the few modes. Winget et al. (1981) firstly put forward that some modes could be trapped in the hydrogen atmosphere, which were called trapped modes. It is effective to study DAV stars, especially for the hydrogen layer mass, by trapped modes. In previous works (e.g. Brassard et al. 1992; C\'orsico et al. 2002; Benvenuto et al. 2002), trapped modes were identified by selecting the minimal period spacings on the period versus consecutive period spacing diagram. Such a method requires a series of pulsation periods observed continuously in $k$, and missing a few modes breaks the period interval line into several segmented ones, which makes it difficult to recognize definitely the trapped modes. According to the asymptotic theory, g-modes with large radial orders follow approximately the equally spacing law on their pulsation periods. C\'orsico et al. (2007) took the deviations from a uniform period spacing to study PG 1159-type star PG0122+200. Bradley \& Winget (1994) also took this method to investigate the trapped mode properties for GD358 and Handler et al. (2002) adopted it to the analysis of the DBV star CBS114. The deviations from the mean period spacing are independent on the continuity in $k$ and only dependent on the relative identification of $k$. Occasionally missing one or two modes do not affect the deviations of other modes. However, this method is dependent on the accurate $l$ identification and the identification of the relative value of $k$.

Trapped modes can be studied on the period versus consecutive period spacing diagram, and can also be studied by doing deviations from the mean period spacing. For asteroseismology, the fittings to observed periods with calculated periods in grid models, namely detailed period to period fittings, are commonly used. Fu et al. (2013) did the asteroseismology study for HS 0507+0434B, including the grid-model calculations (Dolez \& Vauclair 1981) and detailed period to period fittings. Castanheira \& Kepler (2009) did the asteroseismology study for 83 ZZ Ceti stars by detailed period to period fittings. The grid models are generated by White Dwarf Evolution Code (Castanheira \& Kepler 2008). While, with LPCODE, Romero et al. (2012) made fully evolutionary white dwarf models, taking element diffusion into account. They also did the asteroseismology work for DAV stars by doing detailed period to period fittings. In this paper, we are interested in the detailed period to period fitting method. If there is little or no rotational splitting to be observed, we will usually assume the modes of $l$ = 1 modes according to the equally spacing law or a mixture of $l$ = 1 and $l$ = 2 modes. After all, they are assumptions. Thompson et al. (2008) made the spherical degree identifications for G29-38 through limb darkening effect. With their known $l$ modes, we will do the asteroseismology study for G29-38 by grid-model fittings in this paper. Then, we discuss the reliability of asteroseismology work for DAV stars.

\section{The previous works of G29-38}

G29-38 has an extensive range of both spectroscopic and photometric observations. For example, with optical spectrophotometry, Bergeron et al. (1995) calculated the atmospheric parameters ($Teff$ = 11820 K and log $g$ = 8.14), adopting ML2/$\alpha$ = 0.6. By fitting Balmer lines H$\beta$ to H8, Koester et al. (1997) obtained $Teff$ = 11600 K and log $g$ = 8.05. While, Clemens et al. (2000) fitted lines H$\beta$ to H11 in a time-averaged spectrum and gave their best-fitting model with $Teff$ = 11850 K and log $g$ = 8.05. Koester et al. (2005) studied the Ca abundance and gave the atmospheric parameters for G29-38 (WD 2326+049) of $Teff$ = 12100 K and log $g$ = 7.9. In 2009, Koester et al. studied high-resolution UVES/VLT spectra and reported on the new atmospheric parameters of $Teff$ = 11485$\pm$8 K and log $g$ = 8.007$\pm$0.002 (Koester et al. 2009).

Kleinman et al. (1998) made a summary over 1100 hours of time-series photometry including works of two WET runs, a double-site venture between
SAAO and McDonald, and many years of single site observations. They obtained data in 1985 and from 1988 to 1994. Some modes are observed in some years but disappeared in some other years. This phenomenon is common in other DAV stars, such as EC14012-1446 (Handler et al. 2008; Provencal et al. 2012). The stability of eigenmode is another research. Assuming the eigenmodes as $l$ = 1 modes, Kleinman et al. (1998) proposed a radial order identification. Because of assuming all the modes as $l$ = 1, the model fittings do not have a very good result. As said by Kleinman et al. (1998), the critical mismatch of the $l$ = 1 modes were $k$ = 1 and $k$ = 2 modes. The identified eigenmodes by them are shown in the first column in Table 1. While, the second column expresses the mean values of the modes from different observations from 1985 to 1993, which are chosen by Castanheira \& Kepler (2009). Assuming them as $l$ = 1 modes, Castanheira \& Kepler (2009) obtained their best-fitting model ($Teff$ = 11700 K, $M_{*}$ = 0.665 $M_{\odot}$, $M_{He}$ = $10^{-2}$ $M_{*}$, and $M_{H}$ = $10^{-8}$ $M_{*}$). While, assuming them as $l$ = 1, 2 modes, Romero et al. (2012) showed their best-fitting model with $Teff$ = 11471$\pm$60 K, $M_{*}$ = 0.593$\pm$0.012 $M_{\odot}$, $M_{He}$ = 2.39$\times$$10^{-2}$ $M_{*}$, $M_{H}$ = (4.67$\pm$2.83)$\times$$10^{-10}$ $M_{*}$, and log$g$ = 8.01$\pm$0.03. The asteroseismology results are consistent with atmospheric parameters of Koester et al. (2009). However, for the fourteen modes, their results show thirteen $l$ = 2 modes and only one $l$ = 1 mode. It makes us wonder whether there are really so many $l$ = 2 modes observed.

Robinson et al. (1995) firstly introduced a $l$ determination method by studying limb darkening effect. The use of time series at various wavelength, including the UV, allows us to use the wavelength different response of the different $l$ degree to the limb darkening, due to the different geometry of the different $l$ and so help in identifying $l$. With this method, Thompson et al. (2008) made $l$ identifications for G29-38. Their results are shown in the third column in Table 1, which include results of Clemens et al. (2000) and Kotak et al. (2002). The $l$ identifications are shown in the parenthesis. With identified spherical degrees, we try to do the asteroseismology work for G29-38 again.

\begin{table}
\bc
\begin{minipage}[]{100mm}

\caption[]{Detected modes for G29-38.}\end{minipage}
\small
 \begin{tabular}{cccccccccc}
  \hline\noalign{\smallskip}
&$P^{obs1}$& $P^{obs2}$& $P^{obs3}(l)$\\
  \hline\noalign{\smallskip}
&110   &            &               \\
&177   &            &               \\
&237   &218.7       &               \\
&284   &283.9       &284(1)         \\
&355   &363.5       &353(4or3)      \\
&400   &400.5       &               \\
&      &            &431(1)         \\
&500   &496.2       &               \\
&552   &            &               \\
&610   &614.4       &614(1)         \\
&649   &655.1       &655(1)         \\
&678   &            &681(2)         \\
&730   &            &               \\
&771   &770.8       &776(2)         \\
&809   &809.4       &815(1)         \\
&      &            &835(1)         \\
&860   &859.6       &               \\
&894   &894.0       &               \\
&915   &            &920(2)         \\
&      &            &937(1)         \\
&      &1150.5      &               \\
&1147  &            &               \\
&      &1185.6      &               \\
&1240  &1239.9      &               \\
\noalign{\smallskip}\hline
\end{tabular}
\ec
\end{table}

\section{Model calculations and fitting results}

Our white dwarf models are generated by the White Dwarf Evolution Code (WDEC), which was firstly written by Schwarzschild and modified by Kutter \& Savedoff (1969), Lamb \& Van Horn (1975), and Wood (1990). Itoh et al. (1983) introduced the radiative opacities and conductive opacities. The equation of state (eos) are composed of two parts. In the degenerate and ionized core, it takes eos of Lamb (1974). While, in the outer layer, it adopts eos of Saumon et al. (1995). For special C/O mixtures and H/He mixtures, additive volume technique is adopted (Fontaine et al. 1977). For element diffusion, the gravity sedimentation effect results in composition stratified structure. However, the chemical gradient mixes elements in transition regions. DAV stars have experienced a long-time evolution and their shell profile is likely to be or close to equilibrium. So, WDEC adopts equilibrium profiles proposed by Wood (1990) as an approximation of H/He and He/C mixtures. Convective mixing should not reach the H/He transition zone in the temperature range of the ZZ Ceti instability strip except for extremely low mass of the hydrogen layer. Adopting the mixing length theory of B\"{o}hm \& Cassinelli, the mixing length parameter ($\alpha$) is defined as a ratio of mixing length to pressure scale height (B\"{o}hm \& Cassinelli 1971; Tassoul et al. 1990). All the models are results of evolutionary calculations. The initial effective temperature is about 100,000 K and the total number of mesh grids are about 1000. With these models, we numerically solve the full equations of linear and adiabatic oscillation, which find eigenfrequencies one by one through scanning.

Then, we briefly present the grid models we adopt. Four quantities ($Teff$, $M_{*}$, $M_{H}$, and $M_{He}$) are involved in the network. The effective temperature ($Teff$) varies from 11000 K to 12000 K with a step of 50 K. The total stellar mass ($M_{*}$) changes from 0.600 $M_{\odot}$ to 0.800 $M_{\odot}$ with a step of 0.005 $M_{\odot}$. Log($M_{H}$/$M_{*}$) varies from -10 to -4 with a step of 0.5. While, Log($M_{He}$/$M_{*}$) just equals -4, -3.5, -3, -2.5 and -2. The mixing length parameter ($\alpha$) equals 0.6, the same with Bergeron et al. (1995). For the core composition profile, Castanheira \& Kepler (2008) took the homogeneous profile. In order to be closer to the composition profile showed in Romero et al. (2012), we take 20\% carbon on the center of the stellar C/O core, 60\% carbon on the surface of the C/O core, and linear profile between the two ends. Then, the grid models are calculated.

In order to select the best-fitting model, we introduce the commonly used variable,
\begin{equation}
\phi=\phi(M_{*},log(M_{H}/M_{*}),log(M_{He}/M_{*}),T_{eff})
    =\frac{1}{n}\sum(|P^{th}(l)-P^{obs}(l)|).
\end{equation}
\noindent In the equation, $n$ is the number of the observed periods we adopt. $P^{th}$($l$) is the model periods calculated and $P^{obs}$($l$) is the real periods observed. It is worth to say that the absolute difference is in terms of the same spherical degree. The model of smallest $\Phi$ is considered as the best-fitting one.

\begin{figure}[t]
\centering
  \includegraphics[width=0.9\textwidth]{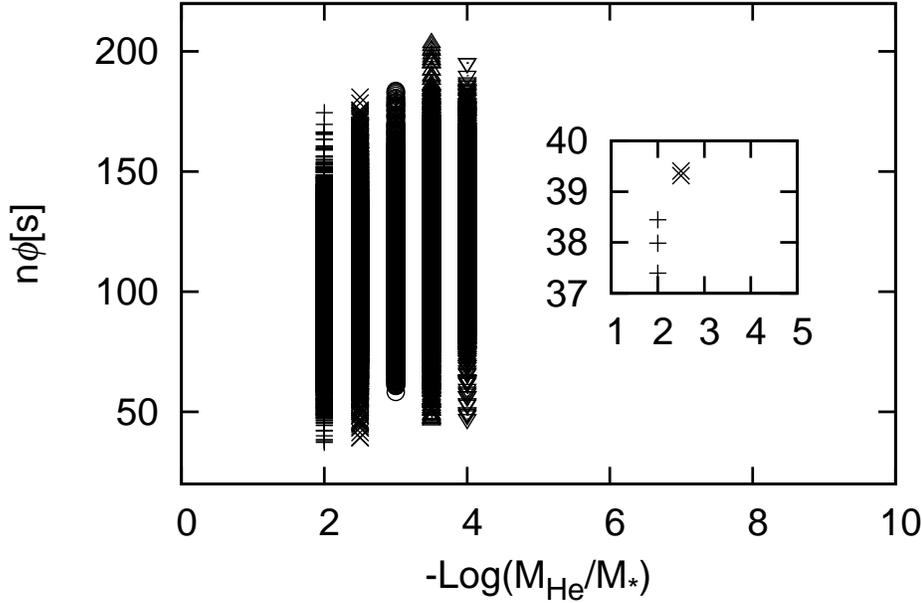}
  \caption{Diagram of selecting best-fitting models. Being from Eq. 1, the ordinate is $n\phi$ and the abscissa is the helium layer mass (-Log($M_{He}$/$M_{*}$)). The diagram shows $n\phi$s for all the grid models and there is a magnified subgraph.}
\label{finger1}
  \end{figure}

\begin{table}
\bc
\begin{minipage}[]{100mm}

\caption[]{Our model results for G29-38.}\end{minipage}
\small
 \begin{tabular}{ccccccccccccccccc}
  \hline\noalign{\smallskip}
&$P^{obs}(l)$& $P^{mod1}(l)$& $\Delta$P(1) & $\phi$(1) & $P^{mod2}(l)$& $\Delta$P(2) & $\phi$(2)\\
  \hline\noalign{\smallskip}
&284(1)    &285.75(1)   &1.75       &3.40      &279.04(1)   &4.96       &3.58      \\
&353(4or3) &353.38(3)   &0.38       &          &357.52(3)   &4.52       &          \\
&431(1)    &429.55(1)   &1.45       &          &433.98(1)   &2.98       &          \\
&614(1)    &606.73(1)   &7.27       &          &609.12(1)   &4.88       &          \\
&655(1)    &644.80(1)   &10.20      &          &658.11(1)   &3.11       &          \\
&681(2)    &683.74(2)   &2.74       &          &680.69(2)   &0.31       &          \\
&776(2)    &776.54(2)   &0.54       &          &777.18(2)   &1.18       &          \\
&815(1)    &810.55(1)   &4.45       &          &815.29(1)   &0.29       &          \\
&835(1)    &838.16(1)   &3.16       &          &847.65(1)   &12.65      &          \\
&920(2)    &917.64(2)   &2.36       &          &918.17(2)   &1.83       &          \\
&937(1)    &940.09(1)   &3.09       &          &934.38(1)   &2.62       &          \\
\noalign{\smallskip}\hline
\end{tabular}
\ec
\end{table}

After all the grid-model fittings to the identified spherical degree modes, the fitting results are obtained and shown in Fig. 1. Being from Eq. 1, the ordinate is $n\phi$ and the abscissa is the helium layer mass (-Log($M_{He}$/$M_{*}$)). At each helium branch, there are grid models of different total stellar masses, different effective temperatures, and different hydrogen layer masses. In addition, there is a magnified subgraph. Therefore, we can clearly see that the minimal $n\phi$s are located in the branches of log($M_{He}/M_{*}$) = -2 and -2.5. For log($M_{He}/M_{*}$) = -2, we take three best-fitting models of $n\phi$ = 37.40, 37.99, and 38.45. They have the same helium layer mass and hydrogen layer mass, and close effective temperatures and total stellar masses ((11900 K, 0.790 $M_{\odot}$), (11750 K, 0.795 $M_{\odot}$), and (11850 K, 0.790 $M_{\odot}$)). We choose the model of $n\phi$ = 37.40 as model1. For log($M_{He}/M_{*}$) = -2.5, we take two best-fitting models of $n\phi$ = 39.32 and 39.41. They have the same helium layer mass and hydrogen layer mass, and close effective temperatures and total stellar masses ((11250 K, 0.780 $M_{\odot}$) and (11150 K, 0.785 $M_{\odot}$)). We choose the model of $n\phi$ = 39.32 as model2. The results of model1 and model2 are shown in Table 2. For the model1, $Teff$ = 11900 K, $M_{*}$ = 0.790 $M_{\odot}$, $M_{He}$ = $10^{-2}$ $M_{*}$, $M_{H}$ = $10^{-4}$ $M_{*}$, log($g$) = 8.30, and $\phi$ = 3.40. While, for the model2, $Teff$ = 11250 K, $M_{*}$ = 0.780 $M_{\odot}$, $M_{He}$ = 3.16$\times$$10^{-3}$ $M_{*}$, $M_{H}$ = 3.16$\times$$10^{-6}$ $M_{*}$, log($g$) = 8.30, and $\phi$ = 3.58. Though 3.40 is less than 3.58, they are basically in the same level. We can not choose that which one is better. What puzzles us is that the fitting error is a little great for the mode of 655 s in our model1 with 644.80 s, for the mode of 835 s in our model2 with 847.65 s, and for the mode of 655.1 s in the model of Romero et al. (2012) with 644.728 s. There is always a relatively great error for one mode at least for each model. For the two models, the gravitational acceleration is larger than the atmospheric results. Anyway, it is a fair choice to fit the known $l$ modes and fitting the known $l$ modes can reduce the blind $l$ fittings. If we do not know the spherical degree of each mode, we can just assume $l$ = 1 or $l$ = 1, 2.

\section{Discussion and conclusions}

In this paper, we focus on the detailed period to period fitting method for asteroseismology. If we assume that all observed eigenmodes are $l$ = 1 modes, the fitting error will be great. For example, $\Phi$ is 7.47 and 7.01 for two best-fitting models respectively for EC14012-1446 in the work of Castanheira \& Kepler (2009). Assuming the same nine observed periods as $l$ = 1, 2 modes, $\Phi$ is 2.54 in the work of Romero et al. (2012) for EC14012-1446. The mean period spacing for different spherical degree is,
\begin{equation}
\bar{\triangle \texttt{P}(l)}=\frac{2\pi^{2}}{\sqrt{l(l+1)}{\int_{0}}^{R}\frac{N}{r}dr}.
\end{equation}
\noindent In the equation, $N$ is buoyancy frequency. According to the equation, the larger the spherical degree, the smaller the mean period spacing. To some extent, if we enlarge the spherical degree, the mean period spacing will decrease and the fitting error will also decrease. For G29-38, assuming $l$ = 1, 2, $\Phi$ is 2.84 in the work of Romero et al. (2012). The effective temperature and gravitational acceleration are consistent with the atmospheric results of Koester et al. (2009). Their white dwarf models are the fully evolutionary ones, which take the element diffusion into account. These models are more realistic for the core composition profile at least, which is really from element burning. However, in their fittings, the fourteen observed modes are thirteen $l$ = 2 modes and only one $l$ = 1 mode. Because there exist so many $l$ = 2 modes, we think that it is necessary to do the asteroseismology work for the star with the modes of identified spherical degree. Thompson et al. (2008) made the $l$ identifications for G29-38 by limb darkening effect. With the known spherical degree modes, we also do the asteroseismology work for G29-38 and the results are different from previous works. The great gravitational acceleration and thick hydrogen layer mass are obtained. The result being different from the previous asteroseismology works is acceptable, because the identification of spherical degree is not the same at least. Actually, the molecular weight gradient for the core composition profile (20\% carbon on the center of the stellar C/O core, 60\% carbon on the surface of the C/O core, and linear profile between the two ends) of our models is smaller than the element burning result of Romero et al. (2012). According to the definition of the buoyancy frequency, the square of buoyancy frequency is proportional to the molecular weight gradient multiplied by the square of gravitational acceleration. In order to obtain the same buoyancy frequency, if we take the core composition profile of element burning result (great molecular weight gradient), the gravitational acceleration will reduce. This may explain why our gravitational acceleration is greater than the atmospheric results.

We do not consider the errors in models themselves temporarily. Just for the identified eigenmodes, such as in Table 1, there are three columns of periods. Some modes appear in one column but not others. There are also modes like 610 s, 614.4 s; 355 s, 363.5 s and so on. All of these will introduce uncertainties for model fittings. The phenomenon is not single. For EC14012-1446, there are nine eigenmodes in Castanheira \& Kepler (2009) and fourteen eigenmodes ($l$ = 1) in Provencal et al. (2012). As we do, taking the known spherical degree modes and doing asteroseismology work is a good choice. It can reduce the error from blind $l$ fittings at least, though there are only eleven modes for G29-38 in the third column in Table 1. For DAV stars, asteroseismology at least depends on the following four factors. Based on factors of only a few modes observed, stability and identification of eigenmodes, identification of spherical degrees, construction of physical and realistic models and so on, detecting the inner structure of DAV stars by asteroseismology needs further development.

Only a few modes being observed. The more the observed modes, the more helpful the model fittings.

The stability and identification problem of eigenmodes. It determines whether we are fitting eigenmodes.

The identification of spherical degrees. Blind spherical degree fittings may lead to some erroneous results.

The construction of physical and realistic white dwarf models. It relates to whether the models themselves are realistic enough to be used for the asteroseismology of white dwarfs.

\section{Acknowledgment}

This work is supported by the Knowledge Innovation Key Program of the Chinese Academy of Sciences under Grant No. KJCX2-YW-T24 and Yunnan Natural Science Foundation (Y1YJ011001). We are very grateful to X. J. Lai, Q. S. Zhang, T. Wu and J. Su for their kindly discussion and suggestions.




\label{lastpage}

\end{document}